\documentclass{iopart}

\usepackage{epsf}

\newcommand{\be}{\begin{equation}} 
\newcommand{\ee}{\end{equation}}
\newcommand{\bea}{\begin{eqnarray}} 
\newcommand{\eea}{\end{eqnarray}}

\newcommand{\gton}{\mathrel{\lower.9ex \hbox{$\stackrel{\displaystyle 
>}{\sim}$}}} 
\newcommand{\lton}{\mathrel{\lower.9ex \hbox{$\stackrel{\displaystyle 
<}{\sim}$}}}  
 
\newcommand{\vp}{{\vec p}}

\begin{document}

\paper{Differential elliptic flow prediction at the LHC from parton transport}

\author{D Molnar}

\address{RIKEN BNL Research Center, Brookhaven National Laboratory, 
Upton, NY 11973, U.S.A.\\
Physics Department, Purdue University, West Lafayette, IN 47907, U.S.A.}
%\ead{graham.douglas@iop.org}
%\begin{abstract}
%\end{abstract}

%Uncomment for PACS numbers title message
%\pacs{00.00, 20.00, 42.10}
% Keywords required only for MST, PB, PMB, PM, JOA, JOB? 
%\vspace{2pc}
%\noindent{\it Keywords}: Article preparation, IOP journals
% Uncomment for Submitted to journal title message
%\submitto{\JPA}
% Comment out if separate title page not required
%\maketitle

\medskip
\noindent
{\bf Introduction.} General physics arguments and calculations
for a class of conformal field theories 
suggest \cite{minetasGyD,minetasSYM} that quantum effects impose a lower 
bound on transport coefficients. For example, the shear viscosity to entropy
density ratio is above a small value $\eta/s \gton 0.1$ 
(``most perfect fluid'' limit).
Dissipative effects can therefore never vanish 
in a finite, expanding system.
On the other hand, ideal (nondissipative) hydrodynamic 
modelling of $Au+Au$ collisions at RHIC ($\sqrt{s_{NN}}\sim 100$ GeV)
is rather successful, leading many to {\em postulate} 
that the hot and dense QCD 
matter created is in fact such a ``most perfect fluid'' 
(at least during the early stages of the RHIC evolution). 
We predict here how differential elliptic flow $v_2(p_T)$ 
changes from RHIC to LHC collision energies 
($Pb+Pb$ at $\sqrt{s}=5.5$~TeV),
{\em if the quark-gluon system created at both RHIC and the LHC
has a ``minimal'' shear viscosity $\eta/s = 1/(4\pi)$}.

\smallskip
\noindent
{\bf Covariant transport theory} 
is a nonequilibrium framework with two main advantages:
i) it has a hydrodynamic limit
(i.e., capable of thermalization); and ii) it is {\em always} causal 
and stable.
In contrast, hydrodynamics (whether ideal, Navier-Stokes, or second-order 
Israel-Stewart theory\cite{IS}) shows instabilities and 
acausal behavior in certain, potentially large, 
regions of the hydrodynamic ``phase space''.

We consider here Lorentz-covariant, on-shell Boltzmann transport theory,
with a $2\to 2$ rate\cite{MPCv2,ZPC}
\bea
p_1^\mu \partial_\mu f_1 &=& S(x, \vp_1) 
+ \frac{1}{\pi} \int\limits_2\!\!\!\!
\int\limits_3\!\!\!\!
\int\limits_4\!\!
\left(f_3 f_4 - f_1 f_2\right)
W_{12\to 34} 
%\nonumber\\ 
%&&\qquad\qquad\qquad\qquad\quad\times 
\  \delta^4(p_1{+}p_2{-}p_3{-}p_4)
\nonumber
\label{Boltzmann_eq}
\eea
The integrals are shorthands
for $\int_i \equiv \int d^3 p_i / (2E_i)$.
For dilute systems, $f$ is the phase space distribution of 
quasi-particles, while the transition probability $W = s (s-4m^2) d\sigma/dt$
is given by the scattering matrix 
element.
Our interest here, on the other hand, is to study the theory 
{\em near its hydrodynamic (local equilibrium) limit}.

Near local equilibrium, the transport evolution 
can be characterized via transport
coefficients of shear and bulk viscosities ($\eta,\zeta$) and heat 
conductivity ($\lambda$) that are determined by the differential cross section.
For the massless dynamics ($\epsilon = 3p$ equation of state) considered here 
$\eta \approx 0.8 T/\sigma_{tr}$, $\zeta = 0$, 
and $\lambda \approx 1.3/\sigma_{tr}$,
$\tau_\pi \approx 1.2 \lambda_{tr}$ \cite{deGroot,IS}
($\sigma_{tr}$ and $\lambda_{tr}$ are the
{\em transport} cross section and mean free path, respectively).

\smallskip
\noindent
{\bf Minimal viscosity and elliptic flow.}
Finite cross sections lead to dissipative effects that reduce elliptic 
flow\cite{pasiv2,ZPCv2}.
For a system near thermal and chemical equilibrium
undergoing longitudinal Bjorken expansion, 
$T \sim \tau^{-1/3}$, $s\approx 4n \sim T^3$, and thus $\eta/s = const$
 requires a growing $\sigma_{tr} \sim \tau^{2/3}$.
With $2\to 2$ processes chemical equilibrium is broken,
therefore
$\sigma_{tr}$ also depends on the density through $\mu/T \sim \ln n$
(because $s = 4(n-\mu/T)$).
We ignore this weak logarithm and take 
$\sigma_{tr}(\tau) = \sigma_{0,tr} (\tau/0.1~fm)^{2/3}$ with 
$\sigma_{0,tr}$ large enough to ensure that most of 
the system is at, or below, the
viscosity bound (thus we somewhat {\em underestimate} viscous effects,
i.e., overestimate $v_2(p_T)$).

For $A+A$ at $b=8$~fm impact parameter we
use the class of initial conditions in \cite{MPCv2}
that has three parameters: parton density $dN/d\eta$,
formation time $\tau_0$, and effective
temperature $T_0$ that sets the momentum scale. 
Because of scalings of the transport solutions \cite{MPCv2},
$v_2(p_T/T_0)$
only depends on two combinations
$\sigma_{tr}\, dN/d\eta \sim A_\perp \tau_0 / \lambda_{tr}$ 
and $\tau_0$. 
This may look worrisome because $dN/d\eta$ at the LHC is uncertain by at least
a factor of two. However,
the ``minimal viscosity''
requirement $T\,\lambda_{tr} \approx 0.5$ {\em fixes} $\sigma_{tr} dN/d\eta$
(e.g., with $dN/d\eta(b{=}0) = 1000$ at RHIC, $\sigma_{0,tr} \approx 2.7$~mb), 
while on dimensional grounds $\tau_0 \sim 1/T_0$.

This means that the {\em main difference between LHC and RHIC is in the typical
momentum scale $T_0$} (gold and lead nuclei are basically
identical),
and therefore to good approximation {\em one expects 
the simple scaling 
$v_2^{LHC}(p_T) \approx v_2^{RHIC}(p_T T_0^{LHC}/T_0^{RHIC})$}.
From gluon saturation physics
we estimate $r\equiv T_0^{LHC}/T_0^{RHIC} \approx 1.3-1.5$ at $b=8$ fm
via Gribov-Levin-Rishkin formula as applied in~\cite{azfarGLR} (we take 
$T_{eff} \sim Q_s \sim \sqrt{\langle p_T^2\rangle}$).

As depicted in Fig.~1, at a given $p_T$ 
the scaling predicts a striking {\em reduction}
of $v_2(p_T)$ at the LHC relative to RHIC. This is the opposite
of both ideal hydrodynamic expectations and 
what was seen going from SPS to RHIC (where $v_2(p_T)$ increased
slightly with energy). Experimental determination of the scaling factor 
$r\equiv Q_s^{LHC}/Q_s^{RHIC}$ 
would provide a further test of gluon saturation models.

\begin{figure}[htpb]
\epsfysize=5.2cm
\epsfbox{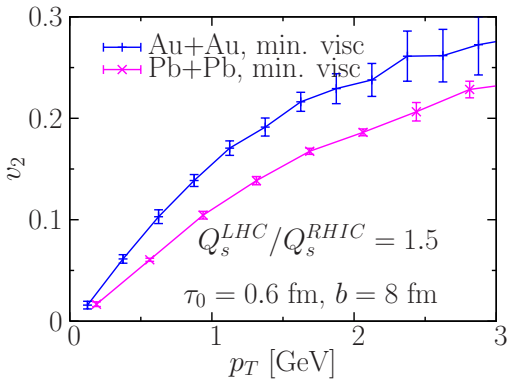}
\caption{Differential elliptic flow at RHIC and the LHC, assuming
a ``minimally viscous'' quark-gluon 
system $\eta/s = 1/(4\pi)$ at both energies.}
\label{fig:1}
\end{figure}

We note that higher momenta at the LHC would also imply
somewhat earlier thermalization 
$\tau_0 \sim 1/T_0$. This is expected to prolong longitudinal Bjorken cooling
at the LHC, changing the scale factor in $v_2(p_T)$ from
$r$ towards $r^{1-1/3} = r^{2/3} \approx 1.2-1.3$.

\medskip
\noindent
{\bf Acknowledgements.} I thank RIKEN, 
Brookhaven National Laboratory and
the US Department of Energy [DE-AC02-98CH10886] for providing facilities
essential for the completion of this work.

\end{document}